\documentclass[graybox]{svmult}

\usepackage{type1cm}        % activate if the above 3 fonts are
\usepackage{makeidx}         % allows index generation
\usepackage{graphicx}        % standard LaTeX graphics tool
\usepackage{multicol}        % used for the two-column index
\usepackage[bottom]{footmisc}% places footnotes at page bottom

\usepackage{newtxtext}       % 
\usepackage{newtxmath}       % selects Times Roman as basic font

\makeindex             % used for the subject index
\begin{document}

\title*{A mixture model approach for clustering bipartite networks}
\author{Isabella Gollini}
\institute{Isabella Gollini \at University College Dublin, Belfield, Dublin, Ireland, \email{isabella.gollini@ucd.ie}}
\maketitle

\abstract*{This chapter investigates the latent structure of bipartite networks via a model-based clustering approach which is able to capture both latent groups of sending nodes and latent variability of the propensity of sending nodes to create links with receiving nodes within each group. This modelling approach is very flexible and can be estimated by using fast inferential approaches such as variational inference. We apply this model to the analysis of a terrorist network in order to identify the main latent groups of terrorists and their latent trait scores based on their attendance to some events.}

\abstract{This chapter investigates the latent structure of bipartite networks via a model-based clustering approach which is able to capture both latent groups of sending nodes and latent variability of the propensity of sending nodes to create links with receiving nodes within each group. This modelling approach is very flexible and can be estimated by using fast inferential approaches such as variational inference. We apply this model to the analysis of a terrorist network in order to identify the main latent groups of terrorists and their latent trait scores based on their attendance to some events.}

\section{Introduction}

In recent years, there has been a growing interest in the analysis of network data. Network models have been successfully applied to many different research areas. We refer to \cite{tow:whi:gol:mur12} for a general overview of the statistical models and methods for networks.

In this chapter we will focus on finding clusters in a particular class of networks that is called bipartite networks. 
Bipartite networks consist of nodes belonging to two disjoint and independent sets, called sending and receiving nodes, such that every edge can only connect a sending node (e.g., actor) to a receiving node (e.g., event).

Latent variable models have been used to model the unobserved group structure of bipartite networks by setting sending nodes as observations and receiving nodes as observed variables (see for example \cite{aitkin2014statistical,ranciati2017identifying}). One important issue limiting the use of classical latent variable approaches, such as latent class analysis \cite{aitkin2017statistical,Bar11} and stochastic blockmodels \cite{now:sni01}, is the assumption of local independence within the groups, that, in presence of a large heterogeneous network, may tend to yield an overestimated number of groups making the results more difficult to interpret and potentially misleading.
Aitkin et al. \cite{aitkin2017statistical} proposed to use different models to overcome the issue of the local dependence assumption including the random Rasch latent class model in which they made use of class and event specific parameters, that are, however, not able to capture the within class behaviour of each actor. %new
Furthermore the computational effort required to estimate the model they propose is significant and this issue makes inference infeasible for large networks.

This chapter concerns the identification of groups in bipartite networks consisting of a set of actors and a set of events through a statistical mixture modelling approach which assumes the existence of a latent trait describing the dependence structure between events within actor groups and therefore capturing the heterogeneity of actors' behaviour within groups. 
This modelling framework allows for: model selection procedures for estimating the number of groups; explanation of the dependence structure of events in each group; description of the behaviour of each actor within each group by quantifying, through the latent trait, the conditional probability that a certain actor belonging to a certain group will attend a certain event. The posterior estimate of the latent trait scores can be visualised so as to interpret the estimated latent traits within each group.
In order to fit the model variational inferential approaches are applied (see \cite{Tip99} and \cite{gol:mur14} for a comparison of estimates given by the variational and other approaches in latent trait models).
The code implemented is included in the {\sf lvm4net} package \cite{lvm4net} for {\sf R} \cite{R}. The rest of this chapter is organised as follow: in Section~\ref{se:MLTA} we describe the model and the inferential approach. In Section~\ref{se:App} we apply the proposed methodology to the Noordin Top terrorist bipartite network \cite{aitkin2017statistical} in which we will aim to identify clusters of terrorists based on their attendance to a series of events in Indonesia from 2001 and 2010. We conclude in Section~\ref{se:Conclusions} with some final remarks.

\section{Model-based Clustering for Bipartite Networks}
\label{se:MLTA}

The relational structure of a bipartite network graph can be described by a random incidence matrix $\mathbf{Y}$ on $N$ sending nodes (i.e. actors), $R$ receiving nodes (i.e. events) and a set of edges $\{ Y_{nr}: n=1,\dots,N; r=1,\dots,R\}$, where: 
\begin{equation*}
Y_{nr} 
=\left\{\begin{matrix}
1, & n \sim r; \\
0, & n \not\sim r.\\
\end{matrix}
\right.
\end{equation*}

To cluster bipartite networks we adapt a flexible model-based clustering approach for categorical data, the mixture of latent trait analyzers (MLTA) model introduced by \cite{gol:mur14}, to the context of bipartite network data. The MLTA model is a mixture model for binary data where observations are not necessarily conditionally independent given the group memberships. In fact, the observations within groups are modelled using a latent trait analysis model and thus dependence is accommodated. The MLTA model generalizes the latent class analysis and latent trait analysis by assuming that a set of $N$ sending nodes can be partitioned into $G$ groups, and the propensity of each actor to create links to the $R$ receiving nodes depends on both the group they belong to and the presence of a $D$ dimensional continuous latent variable $\boldsymbol{\theta}_n$.  

The model assumes that each sending node comes from one of $G$ unobserved groups and defines $\mathbf{z}_{n}=(z_{n1},z_{n2},\ldots,z_{nG})$ as an indicator of the group membership, $z_{ng}=1$ if actor $n$ is from group $g$, with the following distribution:
\begin{equation*}
\mathbf{z}_n\sim\mbox{Multinomial}(1,(\eta_1,\eta_2,\ldots,\eta_G))
\end{equation*}
where $\eta_g$ is the prior probability of a randomly chosen observation coming from group $g$ ($\sum_{g'=1}^{G}\eta_{g'}=1$ and $\eta_{g}\geq 0$ $\forall \, g=1,\ldots,G$). Further, the conditional distribution of $y_{n1},\ldots,y_{nR}$ given that the observation is from group $g$ is assumed to be a latent trait model with parameters $b_{rg}$ and $\mathbf{w}_{rg}$.

Thus, the likelihood of the MLTA model is defined as,
\begin{equation*}
\begin{split}
p(\textbf{y}) 
&= \prod_{n = 1}^N\sum_{g=1}^{G}\eta_g p\left(y_{n1}, \ldots, y_{nR}|z_{ng}=1\right)\\
&=\prod_{n = 1}^N\sum_{g=1}^G  \eta_{g} \int p\left( y_{n1}, \ldots, y_{nR}|\boldsymbol{\theta}_n,z_{ng}=1\right) p(\boldsymbol{\theta}_n)\,d\boldsymbol{\theta}_n
\end{split}
\end{equation*}
where the conditional distribution of given $\boldsymbol{\theta}_n$ and $z_{ng}=1$ is a Bernoulli distribution:
\begin{equation*} \label{mlta.pxy}
\begin{split}
p\left(y_{n1}, \ldots, y_{nR}|\boldsymbol{\theta}_n,z_{ng}=1\right)&=\prod_{r=1}^R p\left( y_{nr}|\boldsymbol{\theta}_n,z_{ng}=1\right)\\
&=\prod_{r=1}^R\left(\pi_{rg}(\boldsymbol{\theta}_n)\right)^{y_{nr}}\left(1-\pi_{rg}(\boldsymbol{\theta}_n)\right)^{1-y_{nr}},
\end{split}
\end{equation*}
and the response function for each group $\pi_{rg}(\boldsymbol{\theta}_n)$ is defined as the following logistic function: % reviewer 2 add Bernoulli
\begin{equation*} \label{mix.pig}
\pi_{rg}(\boldsymbol{\theta}_n)= p\left( x_{nr}=1|\boldsymbol{\theta}_n,z_{ng}=1\right) = \dfrac{1}{1+\exp\left[-( b_{rg}+\mathbf{w}_{rg}^T\boldsymbol{\theta}_n) \right]}, \quad 0\leq \pi_{gr}(\boldsymbol{\theta}_n)\leq 1.  
\end{equation*}
In addition, it is assumed that the $D$-dimensional latent variable $\boldsymbol{\theta}_n\sim \mathcal{N}(\mathbf{0},\mathbf{I}) $.

The attractiveness of receiving node $r$ for sending nodes belonging to group $g$ is modelled by the parameter $b_{rg}$.
The parameter $\mathbf{w}_{rg}$ measures the heterogeneity of the behaviour of sending nodes belonging to group $g$ to connect to the receiving node $r$ (i.e., the heterogeneity of terrorists belonging to the latent group $g$ in attending event $r$); it also accounts for the dependence between receiving nodes.
The vector $\boldsymbol{\theta}_n$ contains the latent variables explaining the propensity of forming links for sending node $n$, i.e., the propensity of terrorist $n$ to attend the events.

We also use a constrained model with common variable-specific slope parameters across groups (i.e. $\mathbf{w}_{rg} = \mathbf{w}_{rg'} = \mathbf{w}_{r}$, where $g \neq g'$): 
\begin{equation*} \label{mix.w.pig}
\pi_{rg}(\boldsymbol{\theta}_n)= \dfrac{1}{1+\exp\left[-( b_{rg}+\mathbf{w}_{r}^T\boldsymbol{\theta}_n) \right]} ,\quad 0\leq \pi_{gr}(\boldsymbol{\theta}_n)\leq 1,
\end{equation*} 

This model is particularly useful to avoid the estimation of too many parameters, especially when the data set is complex, with actors coming from several latent groups and the continuous latent variable having high dimensionality. 

The likelihood of the MLTA model is computationally intractable. For this reason \cite{gol:mur14} proposed to use a double EM algorithm with variational approximation of the likelihood to fit this model, also guaranteeing fast convergence. The main aim of this variational approach is to maximize the Jaakkola \& Jordan \cite{JJ96} lower bound of the likelihood function. This lower bound is a function of auxiliary parameters, called variational parameters, that are optimised to tighten this lower bound. 
The standard errors of the model parameters can be calculated using the jackknife method \cite{Efr81}. For full details of the double EM algorithm we refer to \cite{gol:mur14}.

Since the EM approach is adopted, there is the issue that the results may be affected by the risk of converging to a local maximum instead of the global maximum approximate likelihood. For this reason, it is generally advisable to run the algorithm several times using different initializing values, and select the solution with maximum approximate likelihood. The application of the variational approach makes the estimation procedure much more efficient than most of classical simulation-based estimation methods even when multiple starts are employed.

However, the approximation of the log-likelihood obtained by using the variational approach with the Jaakkola \& Jordan lower bound is always less or equal than the true log-likelihood, so before performing model selection based on the likelihood, like the Bayesian Information Criterion ($\mathrm{BIC}$) \cite{Sch78}, it may be advantageous to get a more accurate estimate the log-likelihood at the last step of the algorithm using Gauss-Hermite quadrature \cite{gol:mur14}. 

\section{Noordin Top Terrorist Network}
\label{se:App}

\begin{figure}
	\centering
	\includegraphics[width=\textwidth]{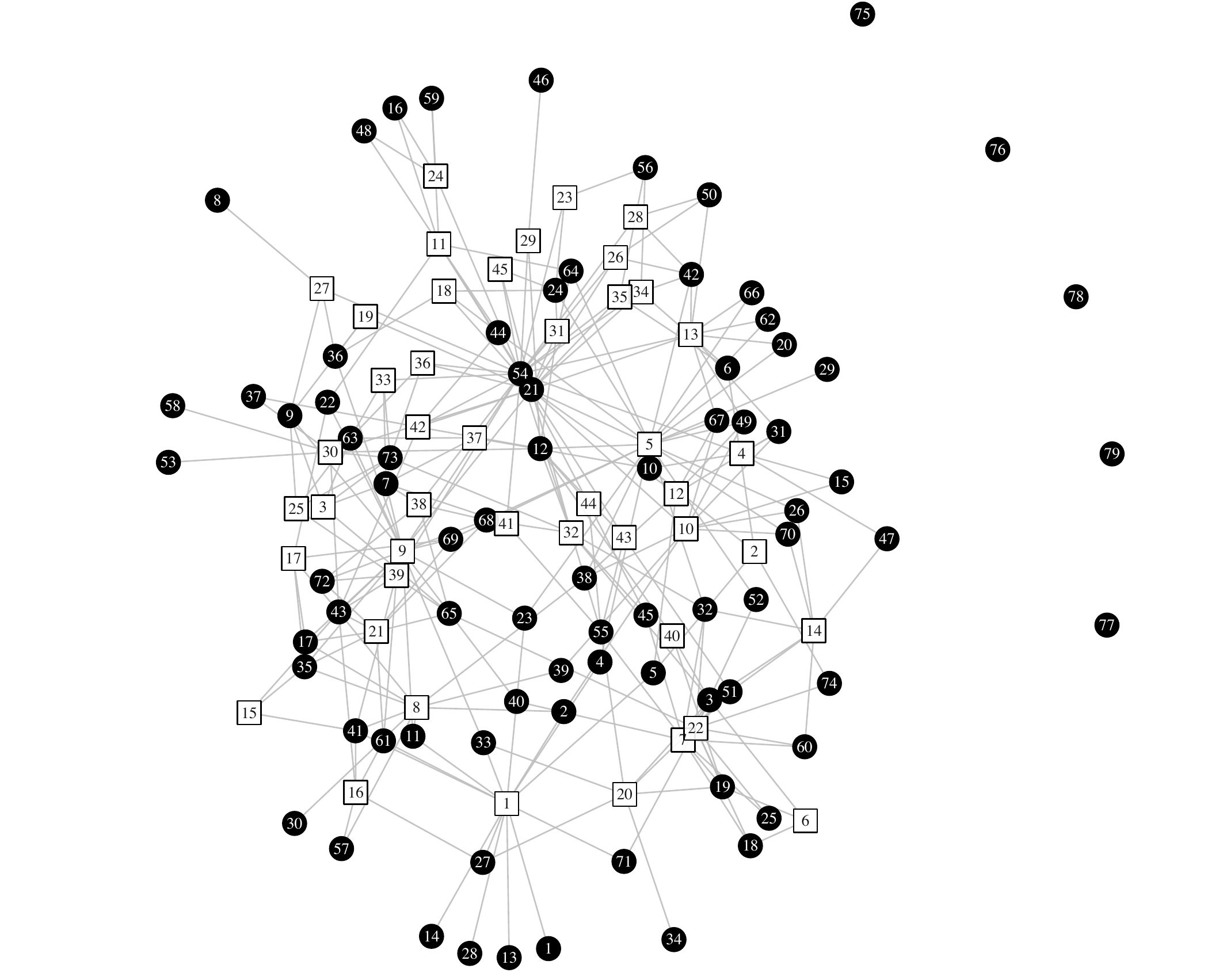}
	\caption{Noordin bipartite graph. Black circles indicate the sending nodes (i.e., terrorists), the white squares indicate the receiving nodes (i.e., events).}  
	\label{fig1}
\end{figure}

The Noordin top terrorist network data \cite{everton2012disrupting} displayed in Figure~\ref{fig1} is a bipartite network oriented around the Malaysian Muslim extremist Noordin Mohammad Top (ID: 54) and his collaborators (the dataset is available in the {\sf manet} package \cite{manet} for {\sf R}). 
The data include relational information on $N = 79$ sending nodes that are individuals belonging to terrorist/insurgent organizations and on $R = 45$ receiving nodes that represent events in Indonesia and nearby areas from 2001 to 2010. The incidence matrix contains links encoding the attendance behaviour of the terrorists to the events. 

\subsection{Statistical Analysis}

We apply the MLTA modelling approach to the Noordin Top Terrorist Network. 
To avoid the issue of getting estimates affected by convergence to a local maximum, we use ten random starts of the algorithm and only the estimates corresponding to  the maximum likelihood value are selected. The model parameters $b_{rg}$ and $\mathbf{w}_{rg}$ are initialized by random generated numbers from a $\mathcal{N}(\mathbf{0},\mathbf{I})$ and the variational parameters are initialized to be equal to 20 in order to reduce the dependence of the final estimates on the initializing values.

The model is fitted on a range of groups, from 2 to 4 and the continuous latent variable takes value $D$ from 0 to 3. For $D = 0$ the MLTA model reduces to a latent class analysis where the observations are assumed to be conditionally independent given the group membership. Model selection is performed on both the unconstrained MLTA and the constrained model with common slope.

The Bayesian Information Criterion ($\mathrm{BIC}$) \cite{Sch78} is used to select the best model, and it is defined as:
\begin{equation*}	\label{mix.bic}
\mathrm{BIC}=-2\ell_{\mathrm{GH}}+ k \log(N),
\end{equation*}
where $\ell_{\mathrm{GH}}$ is the estimate the log-likelihood at the last step of the algorithm obtained by using Gauss-Hermite quadrature, $k$ is the number of free parameters in the model and $N$ is the number of sending nodes. The model with the lower value of $\mathrm{BIC}$ is preferable.

Table~\ref{tab:BIC} shows the BIC values for models with increasing dimensionality. The best model selected is the one with two groups, a one-dimensional latent trait and common slope across groups.

For the best model selected, the values of the mixing proportions are: $\eta_1 = 0.57$ (SE = 0.080) for Group 1, and $\eta_2 = 0.43$ (SE = 0.084) for Group 2.

\begin{table}[ht]
	\centering \caption{BIC results for standard and constrained MLTA models with different number of groups and dimensions.} \label{tab:BIC}
	\begin{tabular}{r|c|cc|cc|cc}
		& $D=0$ & \multicolumn{2}{|c|}{$D=1$} & \multicolumn{2}{|c|}{$D=2$} & \multicolumn{2}{|c}{$D=3$} \\ 
		&     &  &common ${\mathbf{w}_{r}}$ &  &common ${\mathbf{w}_{r}}$&  &common ${\mathbf{w}_{r}}$    \\
		\hline
		$G=2$ & 2062 & 2138 & \textbf{2034} & 2389 & 2096 & 2793 & 2311 \\ 
		$G=3$ & 2157 & 2403 & 2115 & 2876 & 2229 & 3417 & 2434 \\ 
		$G=4$ & 2290 & 2730 & 2249 & 3385 & 2385 & 4419 & 2595 \\ 
		\hline
	\end{tabular}
\end{table}

\subsection{Interpreting the Actor's Behaviour}
\begin{figure}
	\includegraphics[width=\textwidth]{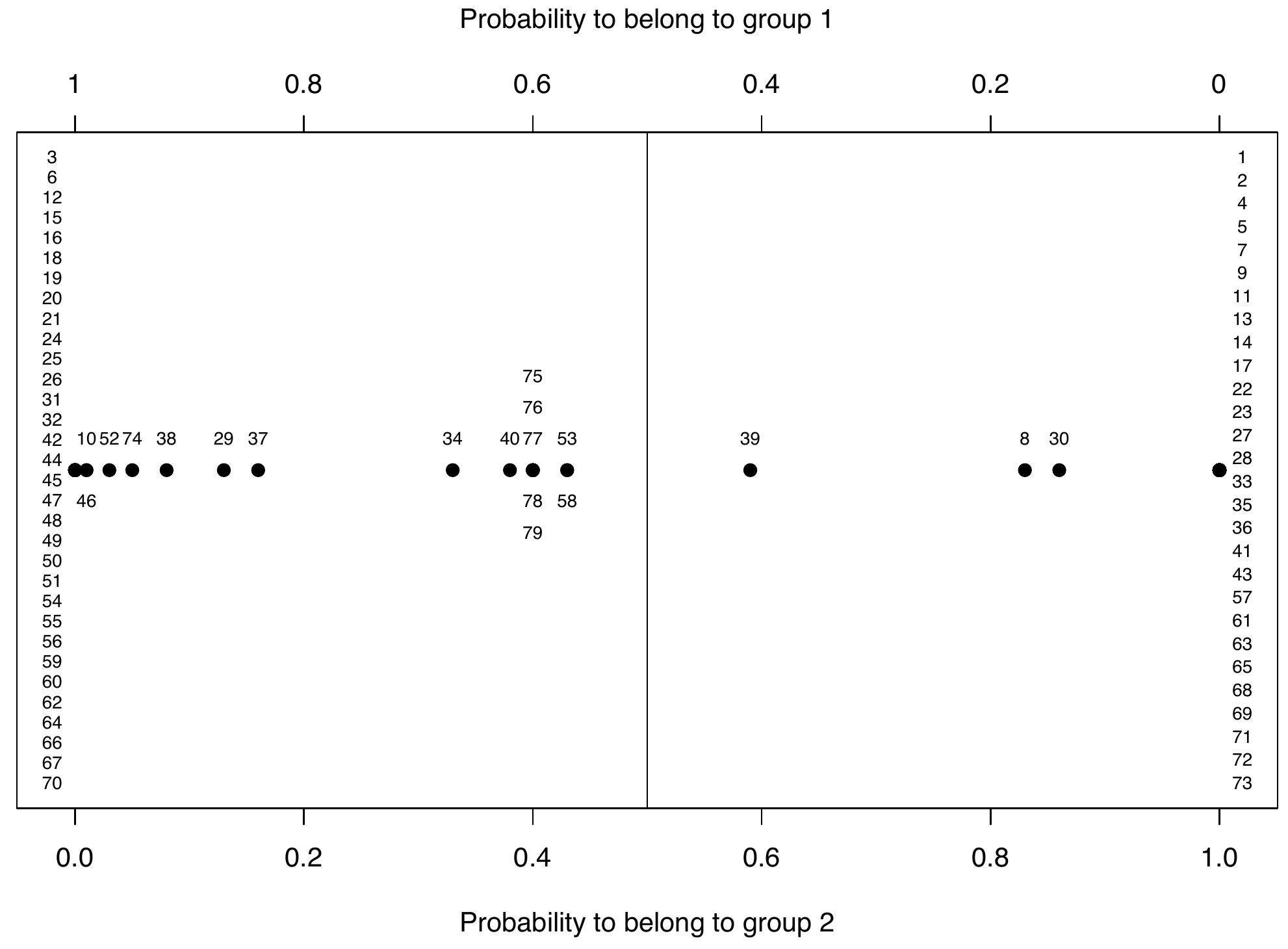}
	\caption{Probability to belong to a group for each terrorist in the best model selected (two groups, a one-dimensional latent trait and common slope across groups).} 
	\label{fig:pz}
\end{figure}

The sending nodes are partitioned into the two groups according to their maximum a posteriori (MAP) probability that they belong to each group. Figure~\ref{fig:pz} shows the posterior probability of each actor to belong to each group. 

Most of the terrorists have been assigned to a particular group with probability very close to 1. In particular, Noordin Top (ID 54), attending 23 events, and Azhari Husin (ID 21), attending 17 events, are allocated together into Group 1 with probability 1. The `lone wolves' (IDs 75, 76, 77, 78, 79), i.e., terrorists who haven't attended any event, have been assigned to Group 1, but the uncertainty associated to their group membership is very large: in fact, their posterior probability to belong to Group 1 is 0.6.

\begin{figure}
	\centering
	\includegraphics[width=0.96\textwidth]{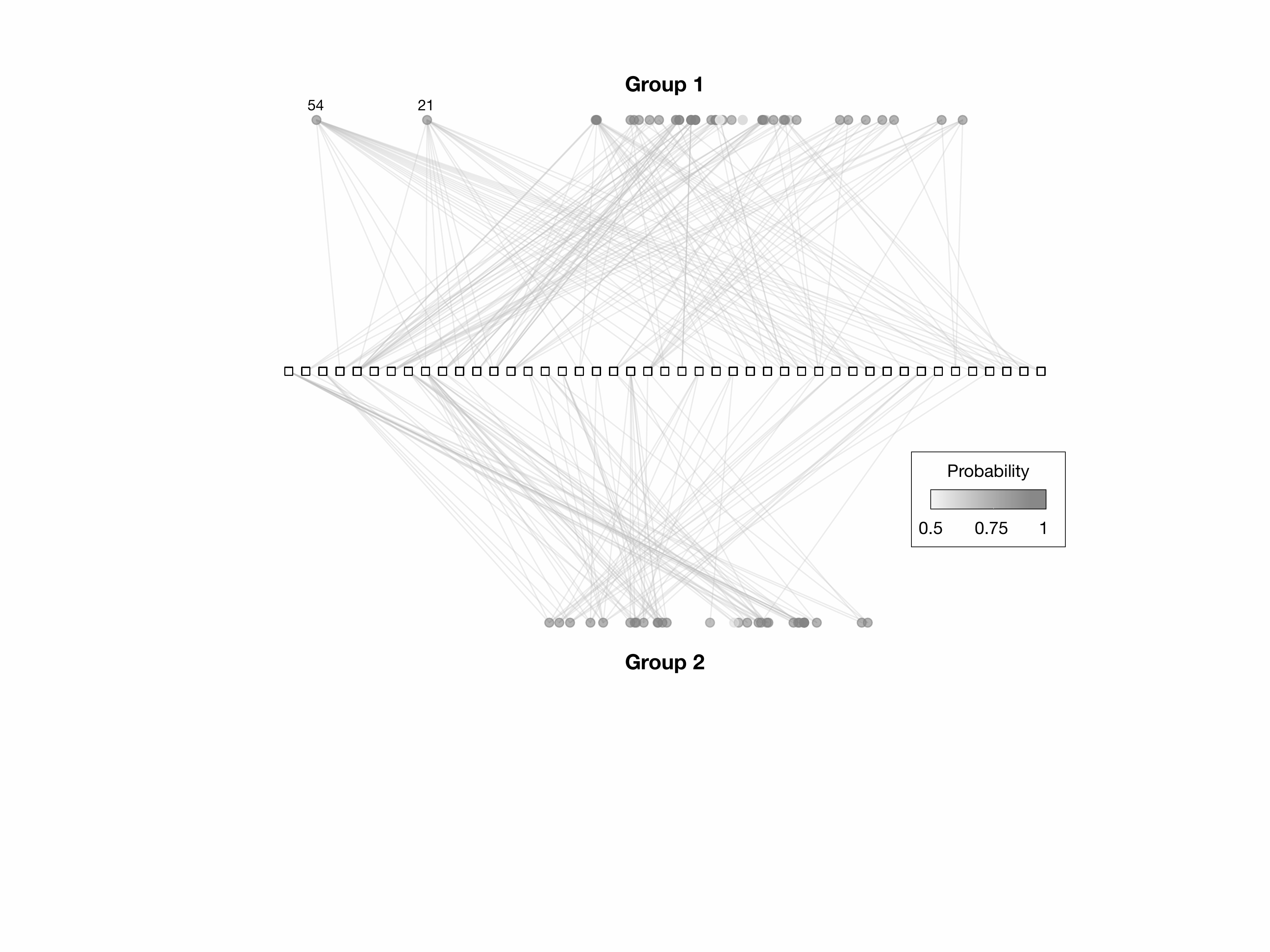}
	\caption{Latent structure of the bipartite network. The terrorists are partitioned into the two groups according to their maximum a posteriori (MAP) probability and plotted according to their latent trait position. The darker the vertex color the higher their MAP of belonging to the group.} 
	\label{fig:map}
\end{figure}

In order to have a deeper understanding of group memberships we can use the information provided by the posterior distribution of the latent trait score $\theta_n$ conditional on the observation belonging to a particular group which can be obtained from the model estimates (see Figure~\ref{fig:map}). 

\begin{figure}
	\centering
	\includegraphics[width=\textwidth]{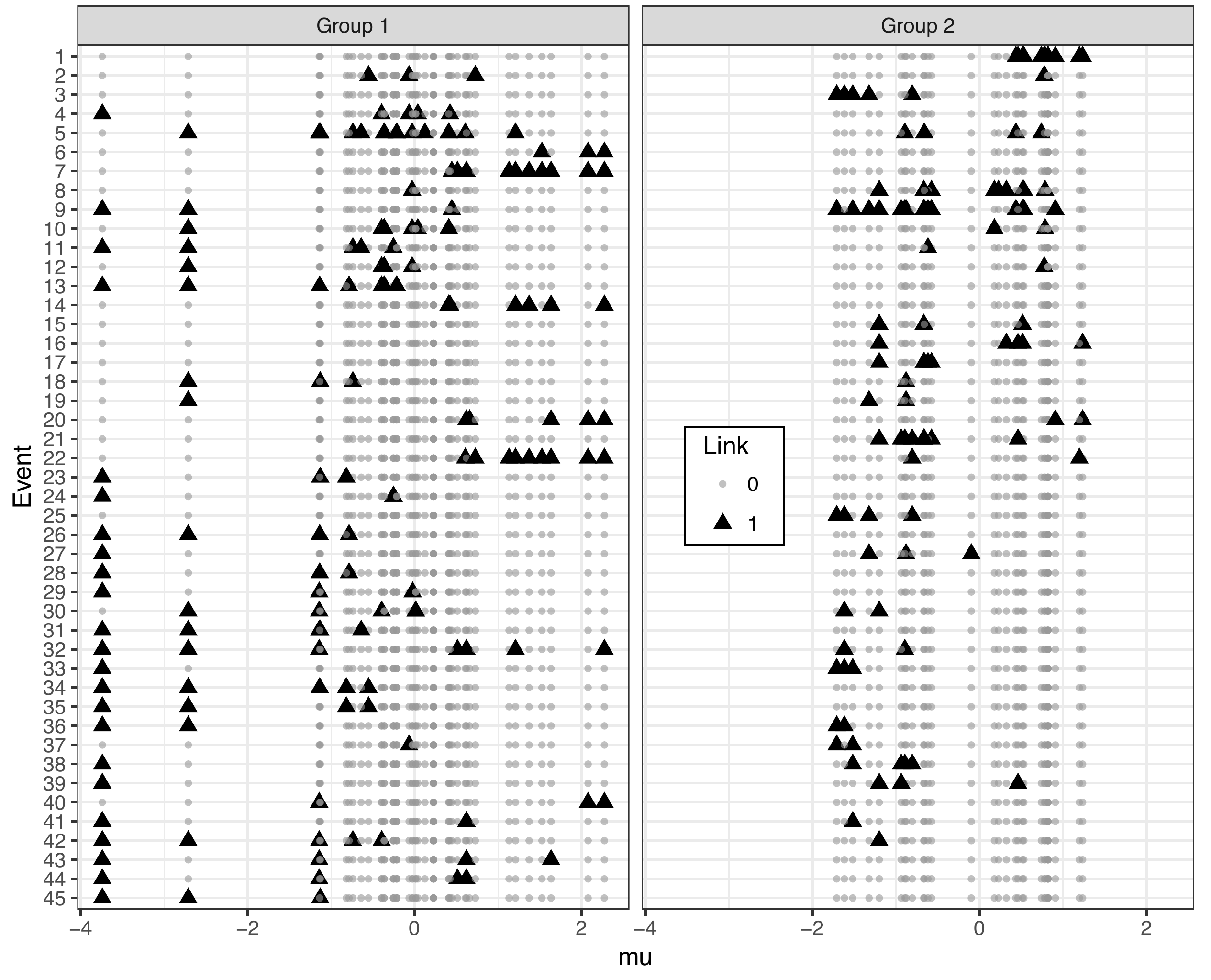}
	\caption{Posterior mean estimate of the latent trait scores $\theta_n$ within each group for each actor and attendance to event.} 
	\label{fig:muEvent}
\end{figure}

The posterior mean estimates of these $\theta_n$ together with the information about event attendance $y_{nr}$ can be used to interpret the latent variables within each group: Figure~\ref{fig:muEvent} allows us to notice that in Group 1 the terrorists with high values went to events 7, 14, and 22 and most of the terrorists with low values went to events 13, 26, 34, 42.
In Group 2 positive values are assigned to those terrorists who attended event 1 (it is also possible to notice that none of the terrorist in Group 1 attended event 1), negative values of the latent trait are associated with terrorists who attended events 2, 9, 25, and 33.

\subsection{Interpreting the Events Attendance}

A measure of the heterogeneity of attending event $r$ within group $g$ is given by the slope value $\mathbf{w}_{rg}$; the larger the value of $\mathbf{w}_{rg}$ the greater the differences in the probabilities of sending a link (going to event) $r$ for actors from group $g$.

\begin{figure}
	\includegraphics[width=\textwidth]{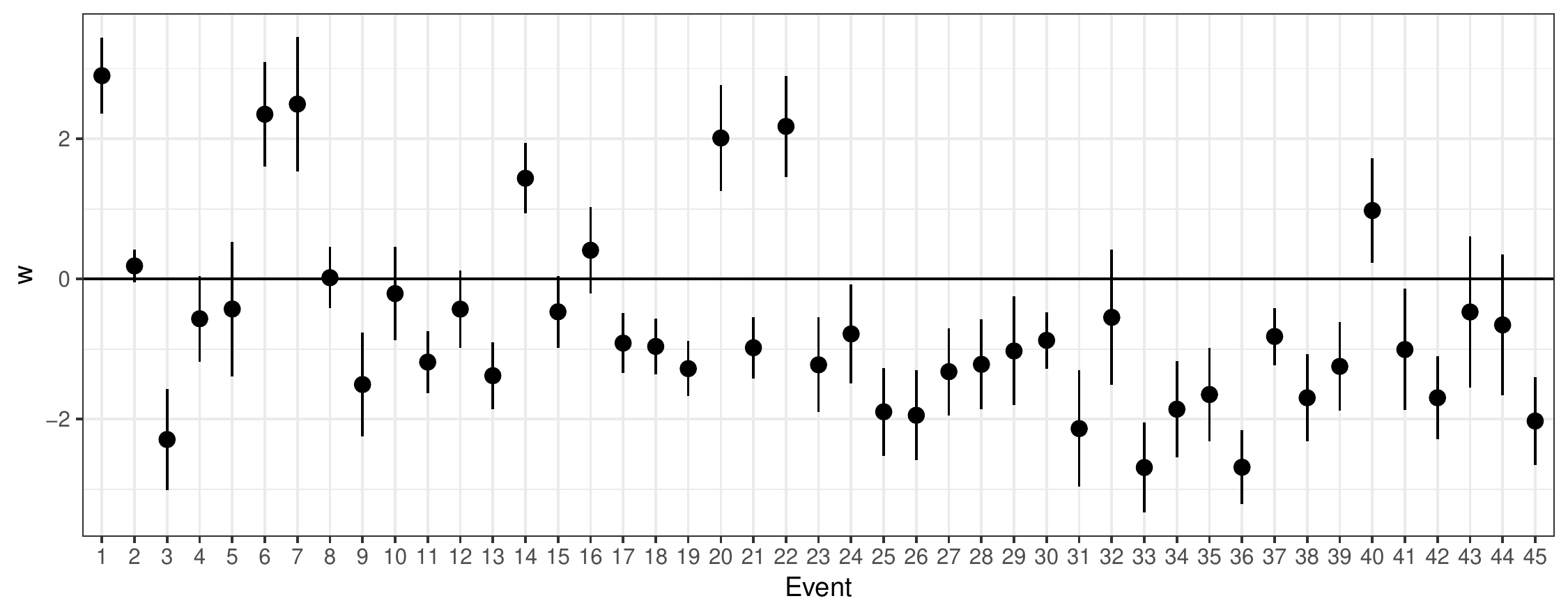}
	\caption{Estimates of slope parameters for each receiving node (event) in the network and associated 95\% confidence interval} 
	\label{fig:w}
\end{figure}

\begin{figure}
	\centering
	\includegraphics[width=.67\textwidth]{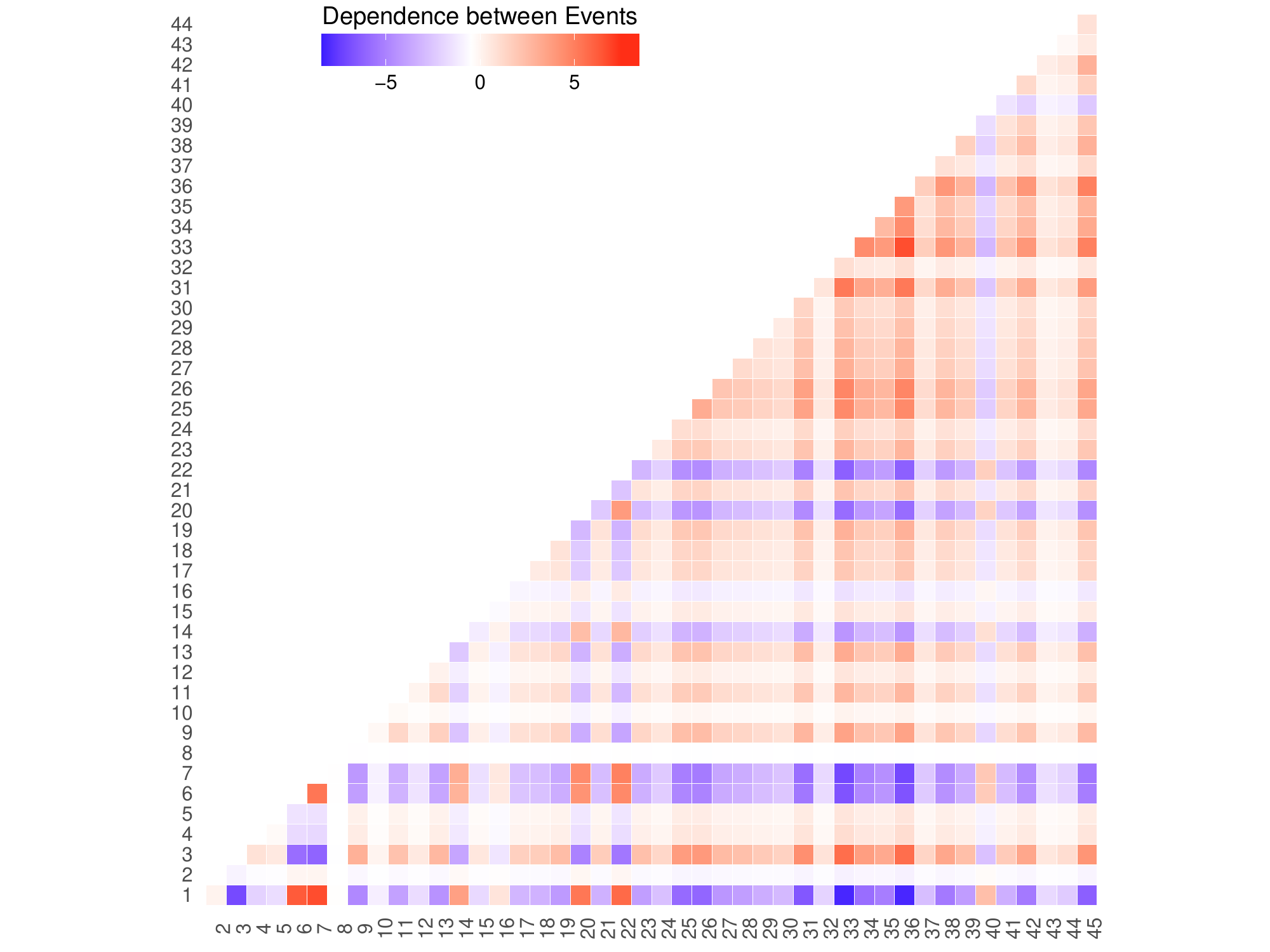}
	\caption{Dependence between events, calculated as  $\mathbf{w}_{r}^T\mathbf{w}_{k}$} 
	\label{fig:wtw}
\end{figure}

The choice of a model with the common slope ($w_{r1} = w_{r2} = w_{r}$) in all groups suggests the latent trait has the same effect in all groups. From Figure~\ref{fig:w} it is possible to notice that most of the slope parameters are non-zero, meaning that the latent trait introduces significant variation within the groups. This indicates that there is considerable variability within the event attendance in the two groups and that some events are positively dependent (i.e., those going/not going to one event will tend to be going/not going in the others events) and other are negatively dependent. The dependence between events $r$ and $k$ in group $g$ is given by $\mathbf{w}_{rg}^T\mathbf{w}_{kg}$, and the results are shown in Figure~\ref{fig:wtw}. Red (blue) squares in the heatmap mean positive (negative) dependence, the darker they are the higher is the dependence between two events. Figure~\ref{fig:wtw} shows that the two set of events $(1, 6, 7)$ and $(3, 33, 36)$ are positively dependent within them and negatively between them.

\begin{figure}
	\includegraphics[width=\textwidth]{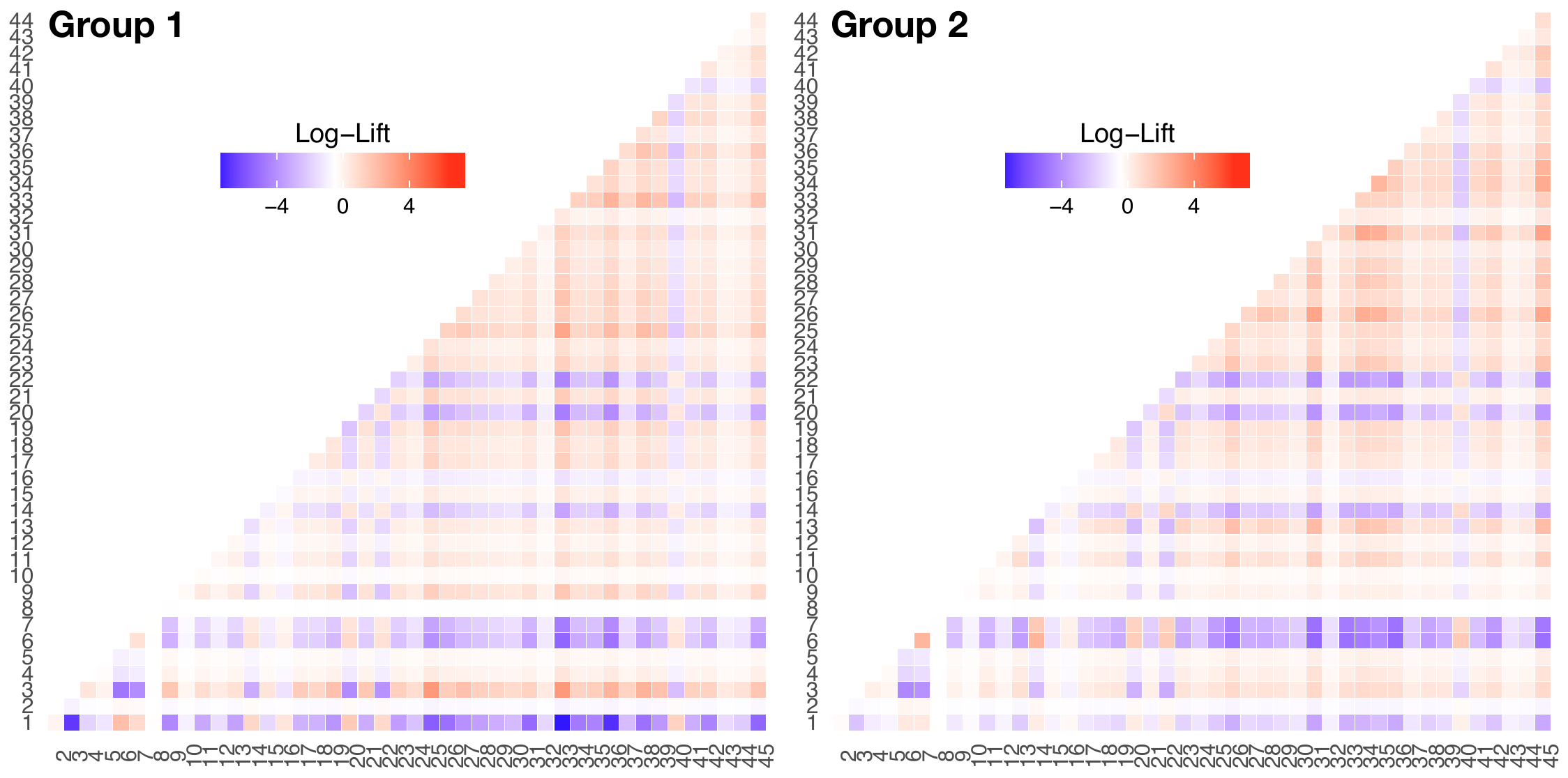}
	\caption{Log-Lift for each pair of receiving nodes (events) of the network} 
	\label{fig:lift}
\end{figure}

The heatmap displayed in Figure~\ref{fig:lift} represents the values of the $\log\{\mathit{lift}\}$ \cite{BMUT97} that can be used to quantify within each group the effect of the dependence on the probability of attending two events compared to the probability of attending two events under an independence model. Mathematically the $\log\{\mathit{lift}\}$ for events $r$ and $k$ for actors belonging to group $g$ is defined as,
\begin{equation*} \label{mix.lift}
\log\{\mathit{lift}\}=\log\left\{\dfrac{\mathbb{P}\left(y_{nr}=1,y_{nk}=1|z_{ng}=1 \right)}{\mathbb{P}\left(y_{nr}=1|z_{ng}=1\right)\mathbb{P}\left(y_{nk}=1|z_{ng}=1 \right)}\right\}
\end{equation*}
where $r=1,2,\ldots,R$ and $r\neq k$.
Two independent events have $\log\{\mathit{lift}\}=0$: the more two events are positively dependent, the higher the value of the $\log\{\mathit{lift}\}$. Lift values that are much less than 0 provide evidence of negative dependence within groups.
Figure~\ref{fig:lift} shows that in Group 1 there is high negative dependence between events 1 and events 3, 33, 36, and in Group 2 there is high positive dependence between the events 27, 32, 34, 35, 45.

\begin{figure}
	\includegraphics[width=\textwidth]{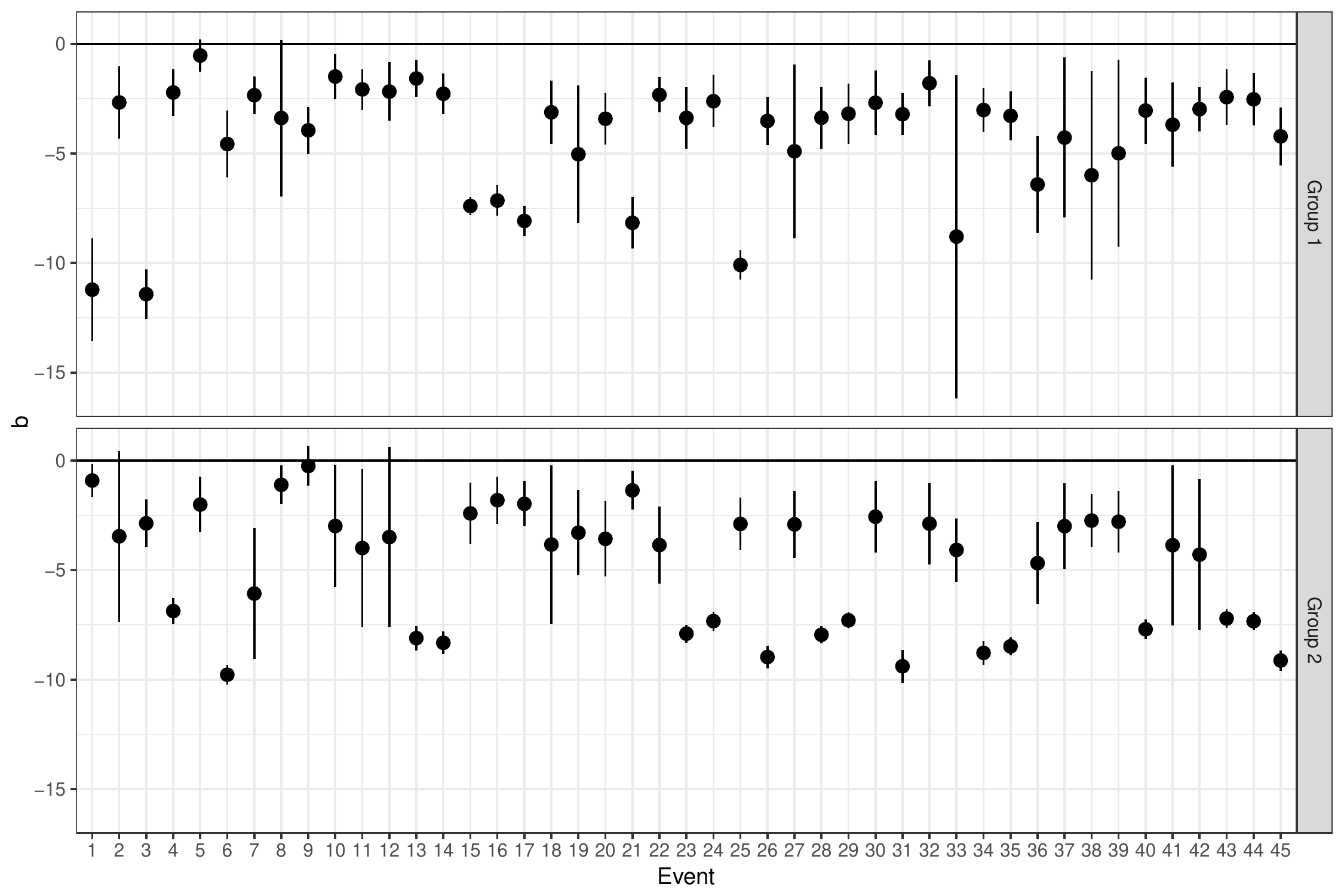}
	\caption{Estimates of the intercept parameters for each receiving node (event) and corresponding 95\% confidence intervals} 
	\label{fig:b}
\end{figure}

The attractiveness of event $r$ for actors belonging to group $g$ is modelled by $b_{rg}$. Figure~\ref{fig:b} shows that most of the values are significantly negative highlighting the sparse structure of the network.

\begin{figure}
	\includegraphics[width=\textwidth]{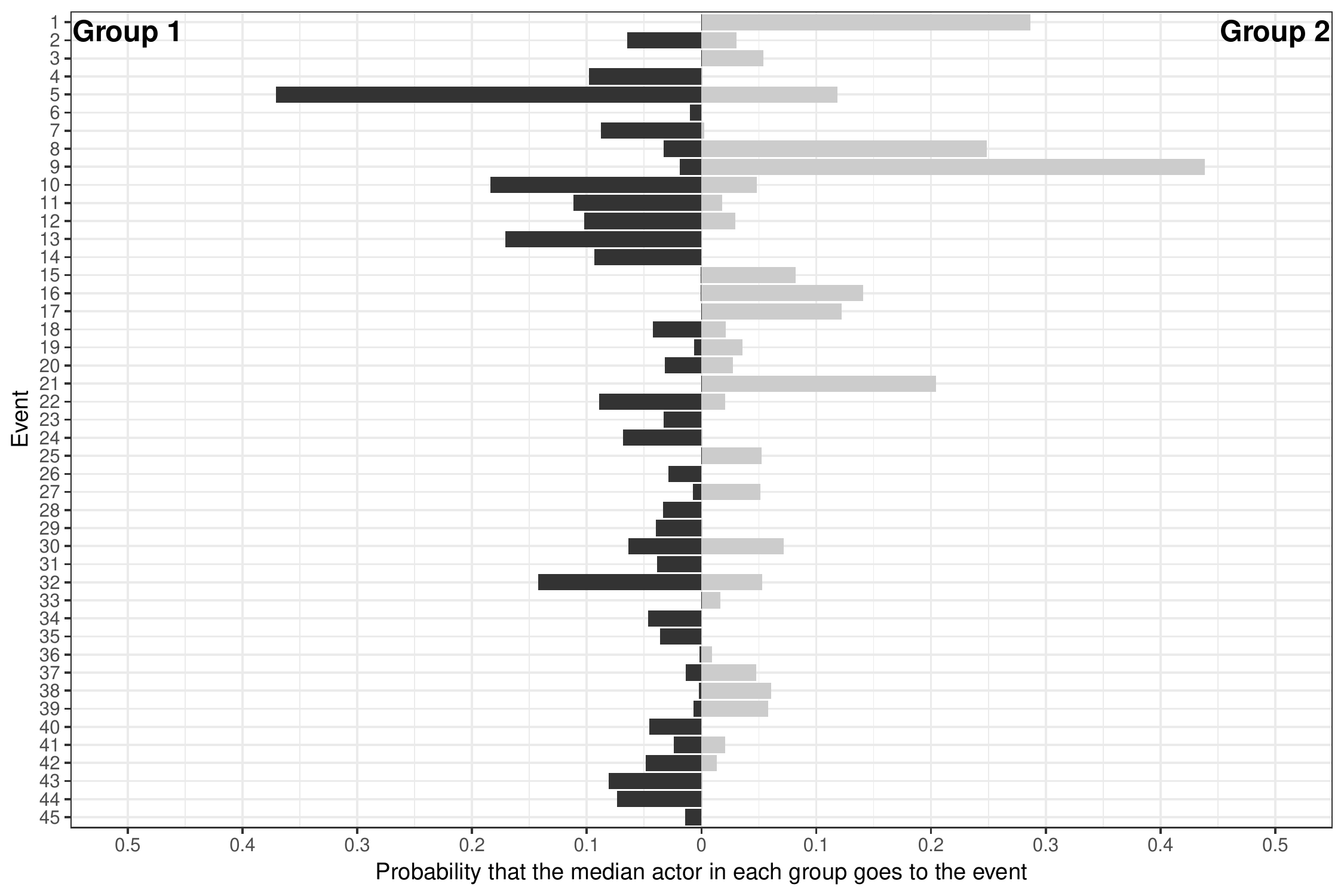}
	\caption{Probability that the median actor in each group goes to the event} 
	\label{fig:pi0g}
\end{figure}

Since $\theta_{n} \sim \mathcal{N}(0,1)$, the probability that the median individual in group $g$ attends events $r$ can be calculated from the attractiveness parameters through the relationship:
\begin{equation*} \label{mix.pi0}
\pi_{rg}(0)=p(x_{nr}=1|\theta_{n} = 0, z_{ng}=1)=\dfrac{1}{1+\exp(-b_{rg})}.
\end{equation*} 
From Figure~\ref{fig:pi0g} it is evident the different behaviour of the actors belonging to the two groups.
Actors in Group 1 have high probability to attend events 7, 13, 14, 43, 44, 45, while those in Group 2 have very low probability to attend these events $(< 10^{-04})$, and none of the terrorists in the data set assigned to Group 2 actually attended those events.
Similarly Actors in Group 2 have high probability to attend events 1, 3, 15, 16, 17, 21, while those in Group 1 have very low probability to attend these events $(< 10^{-04})$, and none of the terrorists in the data set assigned to Group 1 actually attended those events.
Overall the probability that the median actor in each group attends any event is quite low  (the highest probability of $0.438$ for event 9 in Group 2).
This is due to the fact that the number of terrorists attending the same events ranges from a minimum of 3 up to a maximum of 18 out of the total of 79 terrorists.

The proposed methodology is implemented in the {\sf lvm4net} package for R. 

\section{Conclusions}
\label{se:Conclusions}
In this chapter, we have presented an application of a finite mixture model to the clustering of bipartite network data. The modelling framework is particularly flexible and useful for describing the between-group structure using a discrete latent variable and the within-group structure using a continuous latent variable. We have also illustrated how a variational inferential approach can be adopted to estimate the model efficiently. The model has been employed to analyse the relational connectivity patterns of the Noordin Top terrorist network. This has allowed us to find two main groups of terrorists based on their attendance to some events and yield important insights about both terrorists' behaviour within each group and the amount of dependence between events attended by them.

%%%%%%%%%%%%%%%%%%%%%%%% referenc.tex %%%%%%%%%%%%%%%%%%%%%%%%%%%%%%
% sample references
% %
% Use this file as a template for your own input.
%
%%%%%%%%%%%%%%%%%%%%%%%% Springer-Verlag %%%%%%%%%%%%%%%%%%%%%%%%%%
%
% BibTeX users please use
% \bibliographystyle{}
% \bibliography{}
%

\end{document}